\documentclass[twocolumn]{aastex62}

\usepackage{lineno}
\shorttitle{X-ray polarization in blazars}
\shortauthors{Liodakis et al.}

\begin{document}

\title{Prospects for Detecting X-ray Polarization in Blazar Jets}

\correspondingauthor{I. Liodakis}
\email{ilioda@stanford.edu}

\author{Ioannis Liodakis}
\affil{KIPAC, Stanford University, 452 Lomita Mall, Stanford, CA 94305, USA}

\author{Abel L. Peirson}
\affil{KIPAC, Stanford University, 452 Lomita Mall, Stanford, CA 94305, USA}

\author{Roger W. Romani}
\affil{KIPAC, Stanford University, 452 Lomita Mall, Stanford, CA 94305, USA}

\begin{abstract}
X-ray polarization should provide new probes of magnetic field geometry and acceleration physics near the base
of blazar jets, but near-future missions will have limited sensitivity. We thus 
use existing lower energy data and X-ray variability measurements in the context of a basic synchro-Compton
model to predict the X-ray polarization level and the probability of detection success for 
individual sources, listing the most attractive candidates for an {\it IXPE} campaign.
We find that, as expected, several high-peak blazars such as Mrk 421 can be easily measured in
100\,ks exposures. Most low peak sources should only be accessible to triggered campaigns during
bright flares. Surprisingly, a few intermediate peak sources can have anomalously high X-ray polarization
and thus are attractive targets.
\end{abstract}

\section{Introduction}\label{introduc}

The characteristic two peaked spectral energy distribution (SED) of blazar jet emission is understood
to consist of a low energy synchrotron peak and a high energy maximum generally attributed to
Compton emission, although in some models models hadronic processes may also contribute to the high
energy flux \citep{Boettcher2012}. Blazars are often classified by the synchrotron peak frequency
$\rm \nu_{Sy}$, with Low-peak LSP reaching $\rm \nu F_\nu$ maxima in the mm-IR bands, Intermediate ISP sources
peaking in the optical/UV and HSP peaking in the X-rays.
Dramatic variability, on timescales down to minutes in a few cases
\citep{Ackermann2016}, is another hallmark of blazar emission. While radiation-zone models can reproduce
this general emission pattern, many details remain to be explained and the underlying mechanisms of
jet energization and collimation are still a subject of debate \citep{Blandford2018}.

	Polarization can be an important tool for probing the physics of the acceleration zone, especially
in characterizing the magnetic field structures that control the expected shocks and induce synchrotron
radiation. VLBI polarization maps have long been effective at measuring jet fields at pc-scale (e.g., \citealp{Hovatta2012}) while
more recently optical polarization has provided new information on the field orientation and variability
in the unresolved core \citep{Blinov2018}. X-ray polarization,
to be measured by the approved  Imaging X-ray Polarimetry Explorer ({\it IXPE}, \citealp{Weisskopf2018}, launch 2021)
and Enhanced X-ray Timing and Polarization mission ({\it eXTP}, \citealp{eXPT2016}, launch $\sim$2025),
offers new opportunities to probe the jet fields and radiation physics, even closer to the
acceleration site. In particular, polarization can help answer whether leptonic or hadronic process 
dominate in a given band (e.g., \citealp{HoachengZhang2017}).

	However, the sensitivities of the near-future missions are modest and long exposures will
be required, so in light of the variability and limited low energy polarization information one
must choose the expected targets with care. We explore such choice here based on a simple synchro-Compton
model. In section \ref{sec:xray_var} we characterize the X-ray variability of sources observed in optical
polarization monitoring programs, 
in section \ref{sec:xray_predict} we use our model to predict X-ray polarization levels ($\rm\Pi_X$), while
in section \ref{sec:xray_detect} we combine these factors to quantify the success probability of an {\it IXPE}
measurement for reasonable exposure in an untriggered observation, identifying a list of prime targets, and suggesting other X-ray bright sources that can also be of interest if they exhibit strong optical
polarization.
We conclude by discussing new measurements that can improve these predictions and monitoring
campaigns that could make additional sources, and additional classes of polarization behavior, accessible
in the X-ray band.

\section{X-ray variability}\label{sec:xray_var}

\begin{figure}
\resizebox{\hsize}{!}{\includegraphics[scale=1]{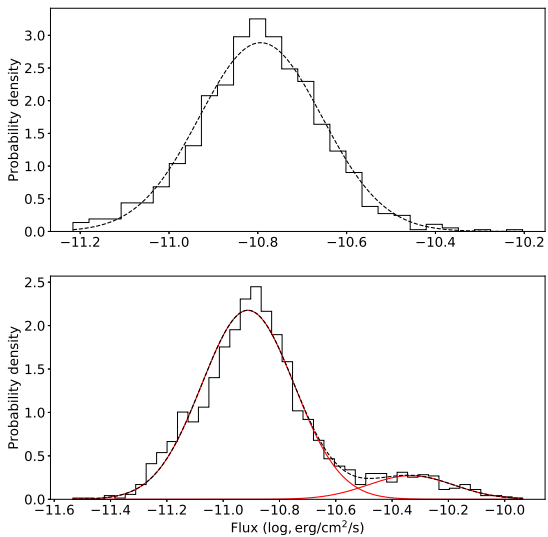} }
 \caption{Example of a unimodal (top panel) and a bimodal (bottom panel) flux distribution in 
the 2-10~keV energy range in log-space. The black dashed line shows the best-fit model. In the bimodal case, 
the red lines show the individual components.}
\label{plt:example_hist}
\end{figure}

	Since X-ray polarization measurements are in general sensitivity limited, source flux variability
plays a key role in the prospects for a secure, 99\% confidence, measurements of a given expected
polarization level. We must therefore characterize the variability of likely targets, the great
majority of which turn out to be sources detected by the {\it Fermi} LAT \citep{Acero2015}. 
We use 2-10\,keV flux measurements from 2005-2017 measured by the
Neil Gehrels {\it Swift} Observatory's (hereafter {\it Swift}) LAT source monitoring 
program\footnote{https://www.swift.psu.edu/monitoring/} \citep{Stroh2013} supplemented by 
2-10~keV fluxes (from 1995-2012) from the RXTE AGN timing and spectral 
database\footnote{We have included all sources in the RXTE database
classified as either BL Lac object (BL Lac) or Flat Spectrum Radio Quasar (FSRQ) with at least 20 observations. https://cass.ucsd.edu/$\sim$rxteagn/} \citep{Rivers2013}. 
\cite{Stroh2013} analyze the individual {\it Swift} observations; we employ their mean 
spectral parameters tabulated for each source to convert epoch count rates to $\rm erg/cm^2/s$ (2-10\,keV) using WebPIMMS. 35 sources (19 LSPs, 2 ISPs, 13 HSPs and one unclassified source\footnote{SED classification from the
3rd {\it Fermi} AGN catalog \cite{Acero2015}, https://www.ssdc.asi.it/fermi3lac/}) have at least
20 observations so that we can attempt a detailed variability analysis. For the remainder, we characterize
their flux variability with a simple mean and standard variation.

	Blazar high energy variability has been modeled as a log-normal distribution 
(e.g., \citealp{Romoli2018,Shah2018}), which may reflect disk-driven fluctuations \citep{Lyubarskii1997}
or variations in the jet particle acceleration \citep{Sinha2018}. This suffices for some of our sources,
but others show wider variability. This may indicate multiple jet states (e.g. quiescent and active
flaring episodes), which can be represented by a double normal (Gaussian mixture) model (in log-space)
\citep{Liodakis2017-IV}. Of course if we have not sampled the full range of a source's variability the 
two log-normals might be subsets of a broader single log-normal. Here, using the historical fluxes,
we represent our flux distribution functions as either single or double log-normal models
without attaching physical significance to the single or double-mode behavior.

The likelihood function for the single Gaussian model is defined as
 \begin{eqnarray}
l_{\rm obs}&=&\frac{1}{\sqrt{2\pi(\sigma_{\rm q}^2+\sigma_{\rm obs}^2)}}
\exp\left[-\frac{(S_{\rm q}-S_{\rm obs})^2}{2(\sigma_{\rm q}^2+\sigma_{\rm obs}^2)}\right],
\end{eqnarray}
where $\rm S_{\rm q}$ and $\rm \sigma_{\rm q}$ are mean and standard deviation of the underlying distribution 
and $\rm S_{\rm obs}$ and $\rm \sigma_{\rm obs}$ are the observed fluxes and their uncertainties (in log-space). For the 
Gaussian mixture the likelihood is defined as
\begin{eqnarray}
l_{\rm obs}&=&\frac{1-f}{\sqrt{2\pi(\sigma_{\rm q}^2+\sigma_{\rm obs}^2)}}\exp\left[-\frac{(S_{\rm q}-S_{\rm obs})^2}{2(\sigma_{\rm q}^2+\sigma_{\rm obs}^2)}\right]\nonumber\\
&+&\frac{f}{\sqrt{2\pi(\sigma_{\rm a}^2+\sigma_{\rm obs}^2)}}\exp\left[-\frac{(S_{\rm a}-S_{\rm obs})^2}{2(\sigma_{\rm a}^2+\sigma_{\rm obs}^2)}\right].
\end{eqnarray}
where we add mean and standard deviation $\rm S_{\rm a}$ and $\rm \sigma_{\rm a}$ for a brighter `active' state
which is realized a fraction $\rm f$ of the observed samples. With such a model, we can draw an arbitrary number
of samples from the modeled distribution. To chose between models for a given source we use the Bayesian 
Information criterion (BIC).  Figure \ref{plt:example_hist} shows examples of best-fit models for 
PKS 0558-504 (top panel) and BL Lacertae (bottom panel). There are 15 sources best described as unimodal,
20 sources prefer a bimodal distribution. LSPs show no preference while HSPs slightly more
commonly match a bimodal distribution (8 versus 5).  The parameters of the best-fit distributions for 
all the sources are given in Table \ref{tab:par_sources}. For sources with $<20$ observations,
we simply record the mean and variance of the log of the flux, which can be used to form a log-normal distribution.

 Measurements are easiest for high polarization $\rm \Pi_X$ sources in bright large $\rm S_{\rm obs}$ states. 
Since both quantities are highly variable, we should test if they correlate, 
as might be expected from e.g. shock-driven flares (e.g., \citealp{Marscher2008}). Of course we lack $\rm \Pi_X$,
so we use the optical polarization ($\rm \Pi_O$) from the Robopol and Steward observatory monitoring programs which have significant
temporal overlap with the {\it Swift} data for 17 sources. We use optical polarization as a tracer of the energetic
electrons at the jet base that may also contribute X-ray synchrotron; radio
polarization can be dominated by downstream emission. While we cannot make meaningful statements about individual
sources, we can check the major source classes by stacking all contemporaneous observations from, say, the HSP.
We find that both the LSP and HSP have a mild positive correlation (Spearman's $\rm \rho\sim0.12$, significance $\rm p-value < 0.05$).
The two ISP showed no correlation. Thus for LSPs and HSPs we draw a flux at
a given level in their cumulative distribution function (CDF, e.g. a flux in the top X\%) and then draw from that source's polarization CDF at the same top X\% level. For the ISP we will assume random uncorrelated draws (see \S \ref{sec:duty_cycle}).
In practice, we find that this makes a small $<20$\% difference to the source detectability, so this assumption is not critical. However it should be tested with future monitoring campaigns.

\section{Expected $\rm \Pi_X$}\label{sec:xray_predict}

	We must use the lower energy (optical) polarization degree $\rm \Pi_O$  to predict the polarization in
the X-ray band. For the HSP and some ISP, the X-rays come from the same (synchrotron) component,
while for the LSP and many ISP, they come from the low energy end of the high energy (here assumed
to be Compton) peak. Particularly interesting are the ISP for which the synchro-Compton transition
occurs within the {\it IXPE} band.
To quantify this connection, we adopt a multizone jet picture \citep{Peirson2018},
where the observed modest $\rm \Pi_O$ are the result of incoherent averaging of $\rm N_{eff}$ effective emission 
zones, each of which radiates with the $\rm \Pi_{max} \approx 70\%$ expected for a uniform field, for a power-law
population of electrons with index $\approx 2$, producing the observed synchrotron spectrum. From observed polarization
levels we typically infer $\rm N_{eff} \approx 30-100$ for the emission cone contributing to the Earth line-of-sight.
In practice, the zones have different angles to the line-of-sight and different characteristic particle energies $\gamma_{max}$
so the number of zones, and thus $\rm \Pi$, becomes a function of the observation frequency 
(e.g. \citealp{Marscher2010-II,Marscher2014}, {\color{blue}Peirson \& Romani 2019, ApJ in prep.}, hereafter PR2019). 

\begin{figure}
\resizebox{\hsize}{!}{\includegraphics[scale=1]{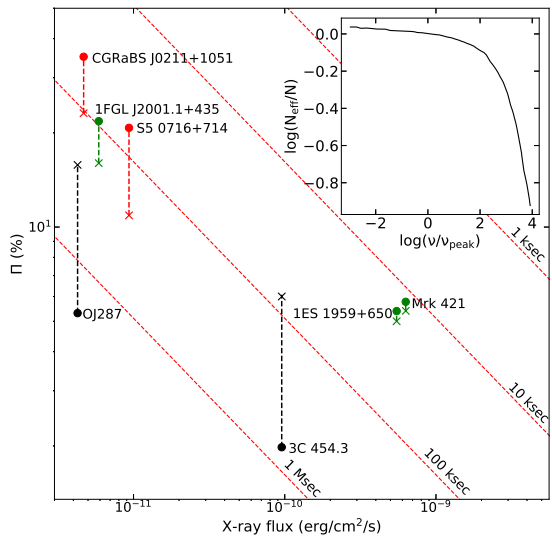} }
\caption{The dashed lines show the shift from observed (optical, ``x'') to predicted (X-ray, ``dots'') $\rm \Pi$
for a few objects in each source class: LSP (black), ISP (red) and HSP (green). The diagonal lines show the {\it IXPE} sensitivity for a source with photon index $\approx2$ in a given exposure time.
Inset shows the frequency dependence of the effective zone number 
using the median Lorentz factor and viewing angle from \cite{Liodakis2018-II}.}
\label{plt:Nzones}
\end{figure}

This multizone picture, with distributions in $\gamma_{max}$ and B field orientation, generally improves the match to observed blazar SEDs over that of a single-zone model. It also means that the $\rm \nu_{Sy}$ for the individual zones vary and so the number of zones $\rm N_{\rm eff}$ contributing half of the integrated flux is a function of
frequency. A computation with $\rm N_{\rm eff}$ related to the peak of the integrated spectrum, assuming typical
jet beaming parameters, $\rm B =0.1~G$ and a uniform squared distribution of $\gamma_{max}$ randomly distributed among
the zones is shown in the inset of Figure \ref{plt:Nzones}. The consequence is $\rm \Pi \approx \Pi_{max}/2\sqrt{N_{\rm eff}}$,
with a small increase from the incoherently averaged half of the flux from the remaining zones (see PR2019 for details). Thus 
$\rm N_{\rm eff}(\nu/\nu_{\rm peak})$ lets us relate the polarization at different frequencies across the
synchrotron component. Note that the $\rm N_{\rm eff}$ decrease and $\Pi$ increase can be dramatic for
$\nu \sim 10^3 \nu_{\rm peak}$; some ISPs can be in this regime.

The behavior in Figure \ref{plt:Nzones}, where the $\gamma_{max}$ range is more important than the effective
Doppler factor variation, is slightly conservative. In some models, such as the shock model of \citet{Marscher2014},
$\gamma_{max}$ may depend on the angle of $B$ to a shock front and hence to the jet axis;
this organized variation further decreases $\rm N_{\rm eff}$ when one is well above the synchrotron
peak.  We find that this effect is only important for $3<\log(\nu/\nu_{peak})<4$, but there the polarization
increase can be as much as an additional $\sim 2\times$; a few ISPs may have synchrotron X-rays from this extreme regime.

For HSP we can directly convert the optical band polarization level to the X-ray band using the 
square root of the ratios of the $\rm N_{\rm eff}(\nu)$. We truncate at $\rm N_{eff}=1$ since our statistical 
estimate breaks down anyway. For some ISP, the $\rm \Pi$ increase can be substantial 
as long as the Compton component contributes weakly at 1-10\,keV. For ISPs, we used the Space Science Data 
Center (SSDC) tools\footnote{https://tools.ssdc.asi.it/SED/} to construct the SED of each source 
and determine whether the X-ray emission is synchrotron dominated. If so, we expect a substantial $\Pi$ increase compared to the optical. 

For LSP (and ISP with hard X-ray spectra) our model assumes that we observe Compton X-ray flux. This will only show polarization if the seed photon population is highly polarized (e.g., synchrotron emission). 
PR2019 find that for isotropic, many-zone scatting in typical jet geometries
the resulting Compton polarization is $0.2-0.36\times$ that of the seed photons. This
does depend on the viewing angle, opening angle, and Lorentz factor of the jet (see PR2019 for details).
However for the typical jet parameters assumed in the present work \citep{Liodakis2018-II} the retained polarization is near
maximal, so we will assume $\Pi_{\rm Comp} = 0.36 \times \Pi_{\rm seed}$. To get the latter, we scale from
$\rm \Pi_O$ using Fig. \ref{plt:Nzones}. For X-ray Compton emission typical seed photons are
in the mm-band, we will assume here $\sim 100$\,GHz, but the dependence on the weighted 
effective seed photon frequency is weak. Note that we are assuming that all seed photons are
synchrotron. If external photons contribute to the seed photon population $\Pi_{\rm Comp}$ will be lower.
This means that our estimates of the LSP polarization may be optimistic. This is useful
since any observed LSP polarization {\it higher} than our estimate indicates that the emission
should be non-Compton in nature (e.g., proton synchrotron).

	With these two effects we predict a $\rm \Pi_X$ for each source (Table \ref{tab:xpol_pred}).
Figure \ref{plt:Nzones} shows the shift from $\rm \Pi_O$ to $\rm \Pi_X$ for a few sources in each class, 
with an inset showing the $\rm N_{\rm eff}$ dependence on frequency.

\section{Blazar Detectability}\label{sec:xray_detect}

\begin{figure*}
\resizebox{\hsize}{!}{\includegraphics[scale=1]{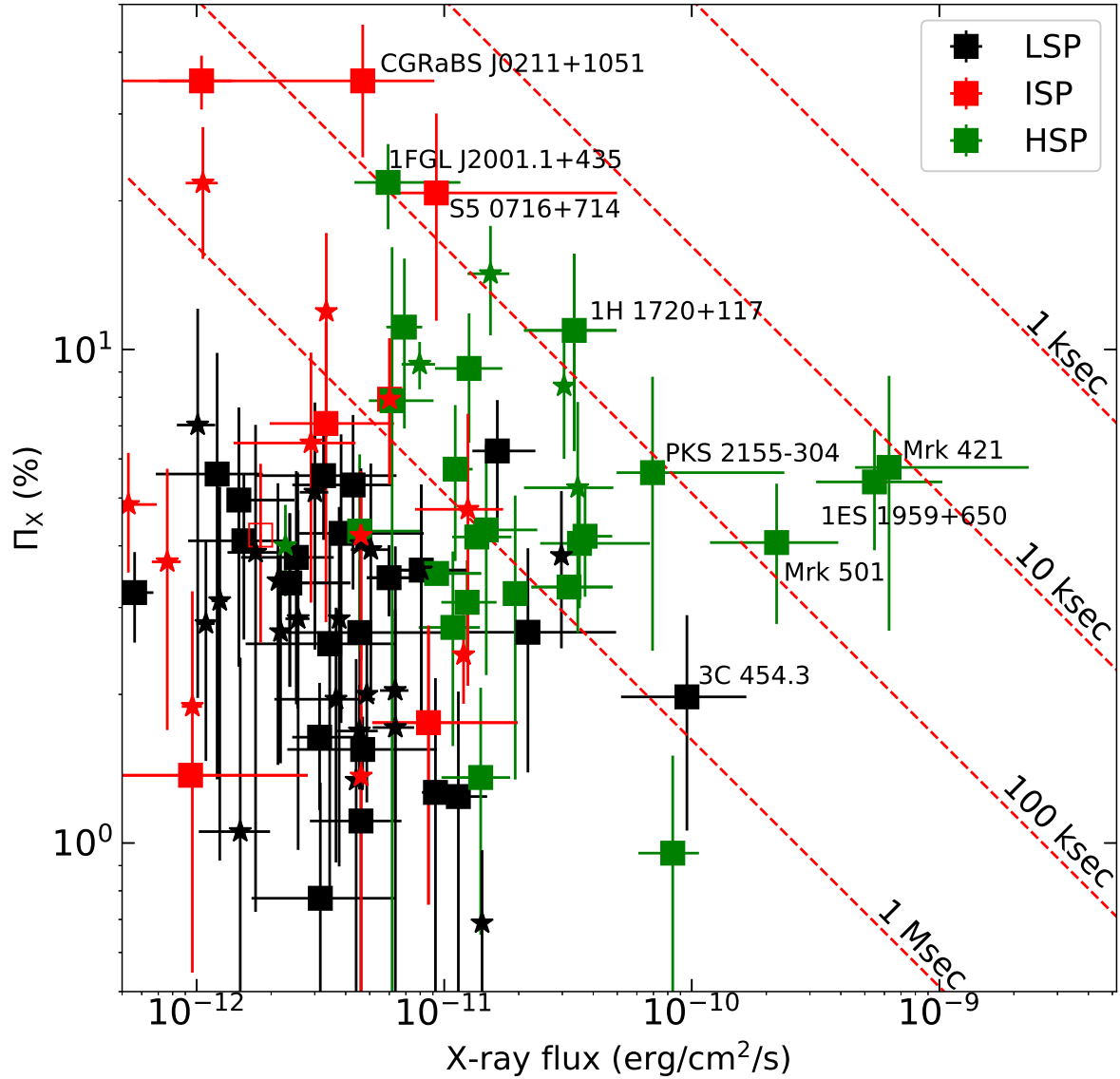} }
\caption{Predicted X-ray polarization degree versus the X-ray flux: LSP (black), ISP (red) and HSP (green).
Lines show the $\rm 1\sigma$ variability range in each quantity. Sources lacking at least 3 observations
each in $\rm S_X$ and $\rm \Pi_O$ are shown by open squares. Stars have $>3$ measurements in one quantity, solid
squares have $>3$ in both.  Red dashed lines show simple estimates for {\it IXPE} sensitivity for a source with photon index $\approx2$
in a given exposure time.}
\label{plt:pred_ixpe}
\end{figure*}

	Our prime target candidates are the sources measured with the RoboPol 
program\footnote{http://robopol.org/} \citep{Pavlidou2014,Blinov2018} and the Steward 
observatory\footnote{http://james.as.arizona.edu/$\sim$psmith/Fermi/} \citep{Smith2009}. 
For some of these we have {\it Swift} and/or RXTE monitoring, and so can construct a detailed
variability model (\S \ref{sec:xray_var}). For the remainder we collect typical fluxes from
the HEARSAC\footnote{https://heasarc.gsfc.nasa.gov/} database. There are 103 
sources with known SED class and at least one measurement in optical polarization and X-ray flux.

	Armed with estimates of the X-ray flux and polarization $\rm \Pi_X$ and their variability
we can make predictions for detectability. We focus on {\it IXPE} as the most imminent facility, whose
sensitivity is estimated using the dedicated online tools\footnote{https://ixpe.msfc.nasa.gov/cgi-aft/w3pimms/w3pimms.pl} \citep{Soffitta2017,ODell2018}. Typical exposures will be $\sim 100\,$ksec, although the longest may approach a Msec. Figure 
\ref{plt:pred_ixpe} shows the X-ray flux and predicted X-ray polarization degree using the median 
and 1$\rm \sigma$ confidence intervals from the PDFs for each source. These PDFs are
estimated following \S \ref{sec:xray_predict}. Note that with our assumed 
correlated fluctuations, HSP and LSP will vary diagonally (UR to LL) within these error bands.
As expected, several HSPs are detectable at 100~ksec while a few (e.g., Mrk 421) might give
significant measurements in 10~ksec, allowing a detailed variability study. Only few LSP sources
are detectable, even with Msec exposures, under typical conditions.  A few 
(e.g., 3C~454.3) are occasionally accessible in shorter time when bright 
and highly polarized.

\subsection{Detectability Duty Cycle}\label{sec:duty_cycle}

	We must consider the substantial flux (\S \ref{sec:xray_var}) and polarization 
 variability (e.g., \citealp{Angelakis2016,Kiehlmann2017}) when predicting the success of
an X-ray polarization search. The uncertainty ranges in Figure \ref{plt:pred_ixpe} already give
some idea of these effects. But some sources vary well outside these ranges, especially
in occasional large $\rm S_X$ flares and less often in polarization increases. Thus we use
distribution function models to characterize the full variability range. For the X-ray variability 
we either use the parameters in Table \ref{tab:par_sources} to construct a flux distribution function or use the mean and standard deviation to define a single log-normal model (see \S \ref{sec:xray_var}). For the optical 
polarization, we use distribution 
functions from the maximum likelihood modeling results of \cite{Angelakis2016} for RoboPol sources; for 
Steward Observatory-monitored sources we use their empirical CDF \citep{Smith2009}.
As noted above for the ISPs we draw randomly for the CDFs, while for LSPs and HSPs we draw an
$\rm S_X$ and then adopt $\rm \Pi_X$ from the same probability level.  We consider only sources with at least three observations in both optical polarization and X-ray 
flux and estimate the joint detection probability (DP) by computing the $\rm MDP_{99}$ in a given 
exposure time and comparing with the predicted $\rm \Pi_X$. We consider
a simulated observation as a detection if $\rm \Pi_X > MDP_{99}$. By repeating this 
calculation $10^4$ times we estimate the fraction of trials a source was detected. 
Dropping the flux-$\Pi$ correlations results in $<20\%$ decrease in the LSP, HSP 
detectability estimates. For the RXTE and {\it Swift} monitored sources we use the average spectral parameters and WebPIMMS to estimate the $\rm MDP_{99}$ from the drawn flux value. For the remaining sources we use a photon index of 1.5 for inverse Compton and 2.5 for synchrotron emitting sources. In any case, assuming different spectral parameters results in only $\sim 5\%$ change in DP. Table \ref{tab:duty_cycle} gives these 
detection probability values for an assumed 100\,ksec {\it IXPE}
exposure. They can be interpreted as the chance of success for a random observation at this
exposure, or as the duty cycle for a triggered (by e.g., flux and/or $\rm \Pi_O$ monitoring) campaign.
Of course, if one wants to measure a particular source, one can obtain more acceptable detection
odds by increasing the exposure duration. While several HSPs have reasonable detection probabilities,
only one ISP (CGRaBs~J0211+1051) and one LSP (3C~454.3) are detected 
at $>$10\% duty cycle. Thus long monitoring campaigns to allow bright trigger thresholds
and/or longer {\it IXPE} exposures will be needed to reliably detect these source classes. It should be noted that source $\rm \Pi_O$ can vary by $2\times$ over a few days so
longer exposures are not strictly `snapshots' as computed here. Intraday variability 
is seen, but is uncommon enough to leave our $\sim$1 day detectability estimates unaffected.

\subsection{Sources without measured optical polarization}

\begin{figure}
\resizebox{\hsize}{!}{\includegraphics[scale=1]{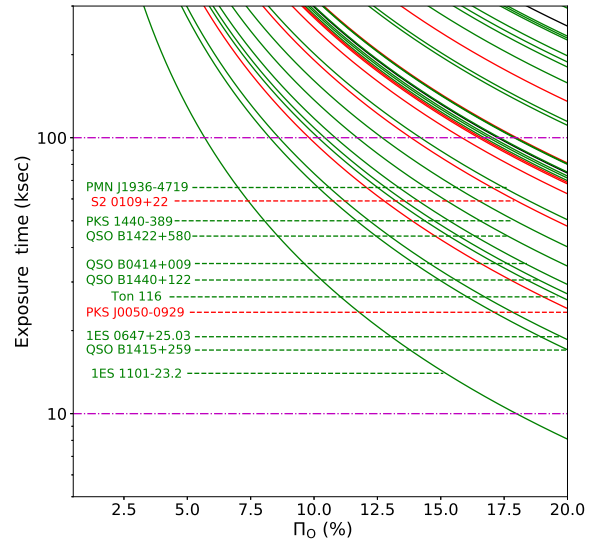} }
 \caption{Required {\it IXPE} exposure time as a function of (presently unknown) optical polarization level.
This $\rm \Pi_O$ has been corrected to an expected $\rm \Pi_X$, using the sources SEDs. The best prospects are
labeled. The sources are colored according to their SEDs: LSPs (black), ISPs (red) and HSPs (green).}
\label{plt:nopol}
\end{figure}

	While many of the best and brightest candidates have been observed in existing optical
polarization campaigns, there are other blazars that might be of interest. For example we find
208 blazars from the BZ catalog \citep{Massaro2015} present in the {\it Swift} master catalog,
97 of which have $S_X>5\times10^{-13}$ and a known spectral $\rm \nu_{Sy}$ class, so that
we can evaluate their observability as a function of the
unknown optical polarization level. With the observed X-ray flux we estimate 
the $\rm MDP_{99}$ (accounting for the different source spectra as in section \ref{sec:duty_cycle}) as a function of exposure time. We convert this to expected optical polarization 
using the relation in Fig. \ref{plt:Nzones}. Figure \ref{plt:nopol} and Table \ref{tab:no_pol} show 
the best prospects from this exercise. Table \ref{tab:no_pol} also lists the 
minimum optical polarization that we would require for $\sim$100\,ksec {\it IXPE} detections.
This suggests that several more HSP and a few ISP are accessible in reasonable 
exposures, although one should obtain reconnaissance $\rm \Pi_O$ measurements first.

\section{Summary}\label{sec:sum}

	We have used the archival SEDs of bright blazars along with observed optical
polarization levels, to predict the expected 1-10\,keV X-ray polarization in a basic 
synchro-self Compton model. This estimate, together with the historical X-ray flux level
lets us evaluate the detectability of X-ray polarization for a given mission sensitivity.
Including the flux and polarization variability as estimated by cumulative distribution
functions modeled from historical data, lets us assess the probability that an exposure of given duration
will achieve success. Equivalently, this gives the duty cycle for observations triggered
by a monitoring campaign to be successful at a given exposure level. We compute these
values for the characteristic {\it IXPE} mission sensitivity, giving a list of top candidate 
sources, useful for planning an observing campaign. 

	Unsurprisingly, HSP dominate the easily detectable sources, but a few ISPs
with X-ray emission well above the synchrotron peak are surprisingly observable. In contrast few
LSP can be accessed, and then only with long exposures. Recalling that our LSP estimate assumes
correlated $\rm S_X/\Pi_O$ variability, and that no external seed photon flux dominates
the up-scatter to the X-ray band, these LSP predictions  should be considered optimistic
for a Synchro-Compton model. However, other emission scenarios (e.g. proton synchrotron)
for the high energy component can produce large $\rm \Pi_X$, so a few LSP observations,
especially when hadronic emission is indicated, would be desirable.

	While our evaluation includes many of the brightest blazars, we have also
identified a set which may be interesting targets, if the typical polarization
level is sufficiently large. Optical reconnaissance to measure these $\rm \Pi_O$ and evaluate
as possible targets for {\it IXPE} and/or {\it eXTP} are strongly encouraged.

\begin{deluxetable*}{lccccccc}
\tablenum{1}
\label{tab:par_sources}
\tablecaption{X-ray flux modeling results.}
\tabletypesize{\scriptsize}
\tablehead{\colhead{Name}& \colhead{Alt. name} & \colhead{SED}  & \colhead{f} & \colhead{$\rm S_q$}  & \colhead{$\rm \sigma_q$} & \colhead{$\rm S_a$}  & \colhead{$\rm \sigma_a$}}
\startdata
J0152+0147 & 1RXS J015240.2+01 & HSP & - & -11.19 & 0.11 & - & - \\ 
J0210-5101 & PKS 0208-512 & LSP & 0.91 & -11.76 & 0.12 & -11.27 & 0.08  \\
J0222+4302 & 3C 66A & HSP & - & -11.16 & 0.06 & - & - \\ 
J0232+2017 & 1ES 0229+200 & HSP & 0.17 & -10.89 & 0.08 & -10.64 & 0.08  \\
J0238+1636 & PKS 0235+164 & LSP & 0.47 & -11.84 & 0.09 & -11.46 & 0.27 \\
J0324+3410 & 1H 0323+342 & HSP & - & -10.85 & 0.14 & - & - \\ 
J0530+1331 & PKS 0528+134 & LSP & - & -11.44 & 0.24 & - & - \\ 
J0539-2839 & PKS 0537-286 & LSP & - & -11.39 & 0.11 & - & - \\
J0559-5026 & PKS 0558-504 & - & - & -10.79 & 0.14 & - & - \\
J0721+7120 & S5 0716+714 & ISP & 0.16 & -11.1 & 0.35 & -10.21 & 0.15  \\
J0831+0429 & PKS 0829+046 & LSP & - & -11.46 & 0.08 & - & - \\
J0830+2410 & QSO B0827+243 & LSP & 0.79 & -11.70 & 0.13 & -11.08 & 0.12  \\
J0841+7053 & 4C 71.07 & LSP & - & -10.82 & 0.09 & - & - \\
J0854+2006 & OJ 287 & LSP & - & -11.37 & 0.18 & - & - \\
J1103-2329 & 1ES 1101-232 & HSP & 0.2 & -10.57 & 0.06 & -10.25 & 0.02  \\
J1104+3812 & Mrk 421 & HSP & 0.44 & -9.52 & 0.46 & -8.94 & 0.28  \\
J1159+2914 & 4C 29.45 & LSP & 0.23 & -11.46 & 0.09 & -11.08 & 0.03  \\
J1221+2813 & W Com & ISP & 0.73 & -12.07 & 0.06 & -11.80 & 0.22  \\
J1229+0203 & 3C 273 & LSP & 0.09 & -9.9 & 0.13 & -9.89 & 0.02  \\
J1256-0547 & 3C 279 & LSP & 0.83 & -11.24 & 0.09 & -10.94 & 0.21  \\
J1408-0752 & 1Jy 1406-076 & LSP & - & -12.25 & 0.09 & - & - \\ 
J1428+4240 & 1H 1430+423 & HSP & 0.52 & -10.64 & 0.2 & -10.08 & 0.12  \\
J1512-0905 & PKS 1510-089 & LSP & - & -11.07 & 0.13 & - & - \\ 
J1555+1111 & PG 1553+113 & HSP & 0.41 & -10.89 & 0.04 & -10.65 & 0.07  \\
J1626-2951 & PKS 1622-297 & LSP & 0.55 & -11.52 & 0.14 & -10.96 & 0.11  \\
J1635+3808 & 1Jy 1633+38 & LSP & - & -11.46 & 0.24 & - & -  \\
J1653+3945 & Mrk 501 & HSP & - & -9.66 & 0.24 & - & - \\ 
J1733-1304 & NRAO 530 & LSP & - & -11.45 & 0.11 & - & - \\
J1959+6508 & 1ES 1959+650 & HSP & 0.79 & -9.93 & 0.22 & -9.27 & 0.25  \\
J2009-4849 & PKS 2005-489 & HSP & 0.51 & -11.01 & 0.09 & -10.03 & 0.35  \\
J2158-3013 & PKS 2155-304 & HSP & 0.61 & -10.62 & 0.25 & -9.94 & 0.25  \\
J2202+4216 & BL Lacertae & LSP & 0.1 & -10.91 & 0.16 & -10.33 & 0.15  \\
J2232+1143 & CTA 102 & LSP & - & -11.02 & 0.08 & - & - \\
J2253+1608 & 3C 454.3 & LSP & 0.93 & -10.69 & 0.1 & -10.01 & 0.22  \\
J2347+5142 & 1ES 2344+514 & HSP & - & -10.47 & 0.23 & - & -  \\  
\enddata
\tablecomments{The X-ray fluxes are all in $\rm erg/cm^2/s$ (log).}
\end{deluxetable*}

\begin{deluxetable*}{lccccccccc}
\tablenum{2}
\label{tab:xpol_pred}
\tablecaption{X-ray flux and polarization.}
\tabletypesize{\scriptsize}
\tablehead{\colhead{Name}& \colhead{Alt. name} & \colhead{Redshift} & \colhead{SED} & \colhead{$\rm \nu_{peak}$}  & \colhead{$\rm S$}  & \colhead{$\rm \sigma_S$} & \colhead{$\rm \Pi_O$} & \colhead{$\rm \sigma_{\Pi_O}$}  & \colhead{$\rm \Pi_X$}}
\startdata
J0017-0512 & CGRaBSJ0017-0512 & 0.227 & LSP & 13.69 & -11.66 & 0.01 & 7.99 & 3.66 & 2.68  \\ 
J0035+5950 & 1ES0033+595 & 0.086 & HSP & 17.12 & -10.5 & 0.26 & 3.1 & 0.01 & 3.3  \\
J0045+2127 & RXJ00453+2127 & -- & HSP & 16.0 & -10.52 & 0.0 & 7.4 & 2.13 & 8.43  \\ 
J0102+5824 & PLCKERC217G124.4 & 0.664 & LSP & 12.94 & -11.52 & 0.06 & 15.9 & 8.27 & 5.14  \\
J0108+0135 & PKS0106+01 & 2.099 & LSP & 13.18 & -11.81 & 0.25 & 12.47 & 4.6 & 4.09  \\ 
J0136+4751 & S40133+47 & 0.859 & LSP & 13.08 & -11.59 & 0.02 & 11.5 & 5.76 & 3.75  \\ 
J0152+0146 & 1RXSJ015240.2+01 & 0.080 & HSP & 15.46 & -11.21 & 0.29 & 6.2 & 6.49 & 7.87  \\
J0211+1051 & CGRaBSJ0211+1051 & 0.200 & ISP & 14.12 & -11.33 & 0.41 & 23.1 & 6.93 & 35.0  \\
J0217+0837 & PLCKERC217G156.1 & 0.085 & LSP & 13.79 & -11.44 & 0.19 & 5.8 & 3.09 & 1.96  \\
J0222+4302 & 3C 66A & 0.340 & HSP & 15.09 & -11.16 & 0.36 & 7.8 & 2.94 & 11.1  \\
\enddata
\tablecomments{The X-ray fluxes are all in $\rm erg/cm^2/s$ (log). Polarization degree is in \%. The table lists sources with $>0.5\%$ X-ray polarization and X-ray flux $\rm >5\times10^{-13} erg/cm^2/s$. The table lists only the first 10 sources. The table is published in its entirety in the machine-readable format. A portion is shown here for guidance regarding its form and content.}
\end{deluxetable*}

\begin{deluxetable*}{lcccc}
\tablenum{3}
\label{tab:duty_cycle}
\tablecaption{Detectability duty cycle.}
\tabletypesize{\scriptsize}
\tablehead{\colhead{Name}& \colhead{Alt. name} & \colhead{SED} & \colhead{$\rm \nu_{peak}$} & \colhead{Det. Prob. (\%)}}
\startdata
J1959+6508 & 1ES 1959+650 & HSP & 16.86 & 72.9 \\
J1725+1152 & 1H 1720+117 & HSP & 16.01 & 60.6 \\
J2001+4352 & 1FGL J2001.1+435 & HSP & 15.21 & 60.3 \\
J1104+3812 & Mrk 421 & HSP & 17.07 & 58.5 \\
J0211+1051 & CGRaBs J0211+1051 & ISP & 14.12 & 49.2 \\
J2158-3013 & PKS 2155-304 & HSP & 15.97 & 42.1 \\
J1653+3945 & Mrk 501 & HSP & 16.12 & 30.7 \\
J0222+4302 & 3C 66A & HSP & 15.09 & 17.0 \\
J1555+1111 & PG 1553+113 & HSP & 15.47 & 14.4 \\
J2253+1608 & 3C 454.3 & LSP & 13.34 & 10.2 \\
J0721+7120 & S5 0716+714 & ISP & 14.6 & 5.6 \\
J1838+4802 & GB6J1838+4802 & HSP & 15.8 & 4.5 \\
J2347+5142 & 1ES 2344+514 & HSP & 15.87 & 3.4 \\
J0958+6533 & S4 0954+658 & LSP & 13.49 & 3.4 \\
J2202+4216 & BL Lac & LSP & 13.61 & 2.5 \\
J1642+3948 & 3C 345 & LSP & 13.23 & 1.8 \\
J1256-0547 & 3C 279 & LSP & 13.11 & 1.8 \\
J0957+5522 & 4C 55.17 & ISP  & 14.23 & 1.5\\
\enddata
\tablecomments{The table is sorted according to detection probability and lists only sources with DP$>1\%$.}
\end{deluxetable*}

\begin{deluxetable*}{lccccccc}
\tablenum{4}
\label{tab:no_pol}
\tablecaption{Sources without optical polarization.}
\tabletypesize{\scriptsize}
\tablehead{\colhead{Name}& \colhead{Alt. name} & \colhead{Redshift} & \colhead{SED} & \colhead{$\rm \nu_{peak}$}  & \colhead{$\rm S$}  & \colhead{$\rm \sigma_S$} & \colhead{$\rm \Pi_{O,min}$}}
\startdata
J0050-0929 & PKS J0050-0929 & 0.635 & ISP & 14.61 & -11.09 & 0.01 & 9.76  \\ 
J0112+2244 & S2 0109+22 & 0.265 & ISP & 14.32 & -11.62 & 0.02 & 13.76  \\ 
J0416+0105 & QSO B0414+009 & 0.287 & HSP & 16.64 & -10.7 & 0.02 & 10.79  \\ 
J0650+2502 & 1ES 0647+250 & 0.203 & HSP & 16.42 & -10.51 & 0.01 & 8.57  \\ 
J1103-2329 & 1ES 1101-23.2  & 0.186 & HSP & 17.19 & -10.07 & 0.01 & 5.66  \\ 
J1243+3627 & Ton 116 & 1.066 & HSP & 16.15 & -10.67 & 0.02 & 10.12  \\ 
J1417+2543 & QSO B1415+259 & 0.236 & HSP & 15.45 & -10.6 & 0.01 & 8.22  \\ 
J1422+5801 & QSO B1422+580 & 0.635 & HSP & 17.72 & -10.73 & 0.03 & 11.65  \\ 
J1442+1200 & QSO B1440+122 & 0.163 & HSP & 16.35 & -10.68 & 0.02 & 10.38  \\ 
J1443-3908 & PKS 1440-389 & 0.065 & HSP & 15.68 & -10.93 & 0.01 & 12.63  \\ 
J1936-4719 & PMN J1936-4719 & 0.265 & HSP & 16.52 & -10.94 & 0.03 & 14.14  \\  
\enddata
\tablecomments{The X-ray fluxes are all in $\rm erg/cm^2/s$ (log). Column $\rm \Pi_{O,min}$ lists the minimum optical polarization degree (\%) required for an {\it IXPE} detection at 100ksec.}
\end{deluxetable*}

\acknowledgements
This work has made use of lightcurves provided by the University of California, San Diego Center for Astrophysics and Space Sciences, X-ray Group (R.E. Rothschild, A.G. Markowitz, E.S. Rivers, and B.A. McKim), obtained at http://cass.ucsd.edu/∼rxteagn/. RoboPol is a collaboration involving the University of Crete, the Foundation for Research and Technology – Hellas, the California Institute of Technology, the Max-Planck Institute for Radioastronomy, the Nicolaus Copernicus University, and the Inter-University Centre for Astronomy and Astrophysics. Data from the Steward Observatory spectropolarimetric monitoring project were used. This program is supported by Fermi Guest Investigator grants NNX08AW56G, NNX09AU10G, NNX12AO93G, and NNX15AU81G and NASA grant NNM17AA26C. This research has made use of data and/or software provided by the High Energy Astrophysics Science Archive Research Center (HEASARC), which is a service of the Astrophysics Science Division at NASA/GSFC and the High Energy Astrophysics Division of the Smithsonian Astrophysical Observatory.
\facilities{{\it Swift}, RXTE, RoboPol, Steward Observatory}

\bibliographystyle{aasjournal}
\bibliography{bibliography} 
\end{document}